\documentclass[aps,prl,twocolumn,showpacs,superscriptaddress,groupedaddress]{revtex4}  % for review and submission
\usepackage{graphicx}  % needed for figures
\usepackage{dcolumn}   % needed for some tables
\usepackage{bm}        % for math
\usepackage{amssymb}   % for math

\def\lumi{7.3}

\def\pythia{{\sc pythia}}
\def\alpgen{{\sc alpgen}}
\def\resbos{{\sc ResBos}}
\def\geant{{\sc geant}}
\def\photos{{\sc photos}}

\newcommand{\phiaco}{\mbox{$\phi_{\rm acop}$}}

\newcommand{\at}{\mbox{$a_T$}}

\newcommand{\zpt}{\mbox{$p_T^{\ell\ell}$}}
\newcommand{\pt}{\mbox{$p_T$}}

\newcommand{\phistarEta}{\mbox{$\phi^{*}_{\eta}$}}
\newcommand{\oneoversigmaphistar}{\mbox{$(1/\sigma)\times (d\sigma/d\phi^{*}_{\eta})$}}
\newcommand{\phistar}{\mbox{$\phi^{*}_{\eta}$}}

\newcommand{\ztt}{\mbox{$Z \rightarrow \tau^-\tau^+$}}

\newcommand{\gtwo}{\mbox{$g_2$}}
\newcommand{\met}{\mbox{$E\kern -0.6em/_{\rm T}$}}

\newcommand{\invfb}{fb\mbox{$^{-1}$}}

\begin{document}
%\setpagewiselinenumbers
%\modulolinenumbers[5]
%\linenumbers

\widetext

% the following line is for submission, including submission to the arXiv!!
\hspace{5.2in} \mbox{Fermilab-Pub-10-403-E}

\title{Precise study of the 
  \boldmath{$Z/\gamma^*$} boson transverse momentum distribution in \boldmath{$p\bar{p}$}
  collisions \\ using a novel technique}
\affiliation{Universidad de Buenos Aires, Buenos Aires, Argentina}
\affiliation{LAFEX, Centro Brasileiro de Pesquisas F{\'\i}sicas, Rio de Janeiro, Brazil}
\affiliation{Universidade do Estado do Rio de Janeiro, Rio de Janeiro, Brazil}
\affiliation{Universidade Federal do ABC, Santo Andr\'e, Brazil}
\affiliation{Instituto de F\'{\i}sica Te\'orica, Universidade Estadual Paulista, S\~ao Paulo, Brazil}
\affiliation{Simon Fraser University, Vancouver, British Columbia, and York University, Toronto, Ontario, Canada}
\affiliation{University of Science and Technology of China, Hefei, People's Republic of China}
\affiliation{Universidad de los Andes, Bogot\'{a}, Colombia}
\affiliation{Charles University, Faculty of Mathematics and Physics, Center for Particle Physics, Prague, Czech Republic}
\affiliation{Czech Technical University in Prague, Prague, Czech Republic}
\affiliation{Center for Particle Physics, Institute of Physics, Academy of Sciences of the Czech Republic, Prague, Czech Republic}
\affiliation{Universidad San Francisco de Quito, Quito, Ecuador}
\affiliation{LPC, Universit\'e Blaise Pascal, CNRS/IN2P3, Clermont, France}
\affiliation{LPSC, Universit\'e Joseph Fourier Grenoble 1, CNRS/IN2P3, Institut National Polytechnique de Grenoble, Grenoble, France}
\affiliation{CPPM, Aix-Marseille Universit\'e, CNRS/IN2P3, Marseille, France}
\affiliation{LAL, Universit\'e Paris-Sud, CNRS/IN2P3, Orsay, France}
\affiliation{LPNHE, Universit\'es Paris VI and VII, CNRS/IN2P3, Paris, France}
\affiliation{CEA, Irfu, SPP, Saclay, France}
\affiliation{IPHC, Universit\'e de Strasbourg, CNRS/IN2P3, Strasbourg, France}
\affiliation{IPNL, Universit\'e Lyon 1, CNRS/IN2P3, Villeurbanne, France and Universit\'e de Lyon, Lyon, France}
\affiliation{III. Physikalisches Institut A, RWTH Aachen University, Aachen, Germany}
\affiliation{Physikalisches Institut, Universit{\"a}t Freiburg, Freiburg, Germany}
\affiliation{II. Physikalisches Institut, Georg-August-Universit{\"a}t G\"ottingen, G\"ottingen, Germany}
\affiliation{Institut f{\"u}r Physik, Universit{\"a}t Mainz, Mainz, Germany}
\affiliation{Ludwig-Maximilians-Universit{\"a}t M{\"u}nchen, M{\"u}nchen, Germany}
\affiliation{Fachbereich Physik, Bergische  Universit{\"a}t Wuppertal, Wuppertal, Germany}
\affiliation{Panjab University, Chandigarh, India}
\affiliation{Delhi University, Delhi, India}
\affiliation{Tata Institute of Fundamental Research, Mumbai, India}
\affiliation{University College Dublin, Dublin, Ireland}
\affiliation{Korea Detector Laboratory, Korea University, Seoul, Korea}
\affiliation{CINVESTAV, Mexico City, Mexico}
\affiliation{FOM-Institute NIKHEF and University of Amsterdam/NIKHEF, Amsterdam, The Netherlands}
\affiliation{Radboud University Nijmegen/NIKHEF, Nijmegen, The Netherlands}
\affiliation{Joint Institute for Nuclear Research, Dubna, Russia}
\affiliation{Institute for Theoretical and Experimental Physics, Moscow, Russia}
\affiliation{Moscow State University, Moscow, Russia}
\affiliation{Institute for High Energy Physics, Protvino, Russia}
\affiliation{Petersburg Nuclear Physics Institute, St. Petersburg, Russia}
\affiliation{Stockholm University, Stockholm and Uppsala University, Uppsala, Sweden }
\affiliation{Lancaster University, Lancaster LA1 4YB, United Kingdom}
\affiliation{Imperial College London, London SW7 2AZ, United Kingdom}
\affiliation{The University of Manchester, Manchester M13 9PL, United Kingdom}
\affiliation{University of Arizona, Tucson, Arizona 85721, USA}
\affiliation{University of California Riverside, Riverside, California 92521, USA}
\affiliation{Florida State University, Tallahassee, Florida 32306, USA}
\affiliation{Fermi National Accelerator Laboratory, Batavia, Illinois 60510, USA}
\affiliation{University of Illinois at Chicago, Chicago, Illinois 60607, USA}
\affiliation{Northern Illinois University, DeKalb, Illinois 60115, USA}
\affiliation{Northwestern University, Evanston, Illinois 60208, USA}
\affiliation{Indiana University, Bloomington, Indiana 47405, USA}
\affiliation{Purdue University Calumet, Hammond, Indiana 46323, USA}
\affiliation{University of Notre Dame, Notre Dame, Indiana 46556, USA}
\affiliation{Iowa State University, Ames, Iowa 50011, USA}
\affiliation{University of Kansas, Lawrence, Kansas 66045, USA}
\affiliation{Kansas State University, Manhattan, Kansas 66506, USA}
\affiliation{Louisiana Tech University, Ruston, Louisiana 71272, USA}
\affiliation{University of Maryland, College Park, Maryland 20742, USA}
\affiliation{Boston University, Boston, Massachusetts 02215, USA}
\affiliation{Northeastern University, Boston, Massachusetts 02115, USA}
\affiliation{University of Michigan, Ann Arbor, Michigan 48109, USA}
\affiliation{Michigan State University, East Lansing, Michigan 48824, USA}
\affiliation{University of Mississippi, University, Mississippi 38677, USA}
\affiliation{University of Nebraska, Lincoln, Nebraska 68588, USA}
\affiliation{Rutgers University, Piscataway, New Jersey 08855, USA}
\affiliation{Princeton University, Princeton, New Jersey 08544, USA}
\affiliation{State University of New York, Buffalo, New York 14260, USA}
\affiliation{Columbia University, New York, New York 10027, USA}
\affiliation{University of Rochester, Rochester, New York 14627, USA}
\affiliation{State University of New York, Stony Brook, New York 11794, USA}
\affiliation{Brookhaven National Laboratory, Upton, New York 11973, USA}
\affiliation{Langston University, Langston, Oklahoma 73050, USA}
\affiliation{University of Oklahoma, Norman, Oklahoma 73019, USA}
\affiliation{Oklahoma State University, Stillwater, Oklahoma 74078, USA}
\affiliation{Brown University, Providence, Rhode Island 02912, USA}
\affiliation{University of Texas, Arlington, Texas 76019, USA}
\affiliation{Southern Methodist University, Dallas, Texas 75275, USA}
\affiliation{Rice University, Houston, Texas 77005, USA}
\affiliation{University of Virginia, Charlottesville, Virginia 22901, USA}
\affiliation{University of Washington, Seattle, Washington 98195, USA}
\author{V.M.~Abazov} \affiliation{Joint Institute for Nuclear Research, Dubna, Russia}
\author{B.~Abbott} \affiliation{University of Oklahoma, Norman, Oklahoma 73019, USA}
\author{M.~Abolins} \affiliation{Michigan State University, East Lansing, Michigan 48824, USA}
\author{B.S.~Acharya} \affiliation{Tata Institute of Fundamental Research, Mumbai, India}
\author{M.~Adams} \affiliation{University of Illinois at Chicago, Chicago, Illinois 60607, USA}
\author{T.~Adams} \affiliation{Florida State University, Tallahassee, Florida 32306, USA}
\author{G.D.~Alexeev} \affiliation{Joint Institute for Nuclear Research, Dubna, Russia}
\author{G.~Alkhazov} \affiliation{Petersburg Nuclear Physics Institute, St. Petersburg, Russia}
\author{A.~Alton$^{a}$} \affiliation{University of Michigan, Ann Arbor, Michigan 48109, USA}
\author{G.~Alverson} \affiliation{Northeastern University, Boston, Massachusetts 02115, USA}
% ...
%
% visitor_addresses.tex                        24 August 2010
%  available symbols are:
%  $\ast, \dag, \ddag, \S, \P, $\|$, $\ast\ast$, \dag\dag, \ddag\ddag ,\#
%
\collaboration{The D0 Collaboration\footnote{with visitors from
%{alton}
$^{a}$Augustana College, Sioux Falls, SD, USA,
%{burdin}
$^{b}$The University of Liverpool, Liverpool, UK,
%{haas,partridge}
$^{c}$SLAC, Menlo Park, CA, USA,
%{juste}
$^{d}$ICREA/IFAE, Barcelona, Spain,
%{luna-garcia}
$^{e}$Centro de Investigacion en Computacion - IPN, Mexico City, Mexico,
%{podesta-lerma}
$^{f}$ECFM, Universidad Autonoma de Sinaloa, Culiac\'an, Mexico,
and 
%{weber}
$^{g}$Universit{\"a}t Bern, Bern, Switzerland.%
%{hooper}
%$^{?}$%Visitor from Bradley University, Peoria, IL, USA.
%{kozminski
%$^{?}$}%Visitor from Lewis University, Romeoville, IL, USA.
%{deceased}
%$^{\ddag}$%Deceased.
}} \noaffiliation
\vskip 0.25cm
       % D0 authors (remove the first 3 lines
                             % of this file prior to submission, they
                             % contain a time stamp for the authorlist)
                             % (includes institutions and visitors)
\date{October 1, 2010}

\begin{abstract}
  Using  \lumi~\invfb\ of $p\bar{p}$
  collisions collected by the D0 detector at the Fermilab Tevatron, 
  we measure the distribution of the variable \phistar, which probes the same physical effects as the $Z/\gamma^*$ boson transverse momentum,
  but is less susceptible to the effects of experimental resolution and efficiency.
%The data are corrected for detector effects and presented in bins of boson rapidity.
A QCD prediction is found to describe the general features of the
\phistar\ distribution, but is unable to describe its detailed
shape or dependence on boson rapidity.
A prediction that includes a broadening of
transverse momentum for small values of the parton
momentum fraction is strongly disfavored.
\end{abstract}

\maketitle

\clearpage
%\section{\label{sec:level1}First-level heading}
% sections are not used for PRL papers

$Z/\gamma^*$ bosons are produced at hadron colliders via quark-antiquark annihilation.
Their decays to $e^+ e^-$ and $\mu^+ \mu^-$
can be detected with little background and the phenomenology is
simplified by the absence of color flow between the
initial and final states, thus providing an excellent
testing ground for QCD predictions.
%Due to the colorless and low background final state, 
%$Z/\gamma^*$ boson decays into electron and muon pairs provide an excellent
%testing ground for QCD predictions at hadron colliders.
%Understanding the transverse momentum, \zpt, distribution
%is of direct interest to Higgs boson and new physics searches 
%at hadron colliders.
Resummation techniques~\cite{CSS1985} allow calculations of the
distribution of $Z/\gamma^*$ boson transverse momentum, \zpt, within the framework
of perturbative QCD, even at relatively low \zpt\ (e.g., \zpt~$<$~30~GeV).
However, additional non-perturbative form factors must be determined in global fits to experimental 
data~\cite{BLNY2003}.
An increase of  these form factors for $x < 10^{-2}$,
where $x$ is the parton momentum fraction, was
suggested~\cite{Nadolsky-smallx-2001} to improve the
description of hadron production observed in deep 
inelastic electron-proton scattering at HERA.
Since vector boson production corresponds
typically to parton $x < 10^{-2}$ at the LHC, 
these modified  form factors would lead to a broadening of the expected vector
boson  transverse momentum distributions~\cite{BergeSmallx-2005}. 
This ``small-$x$ broadening'' would influence the measurement of
the $W$ boson mass as well as searches for Higgs
bosons and physics beyond the standard model at the LHC. 
It is important to study quantitatively such $x$-dependencies at the Tevatron, where
they can be probed using the dependence of
the \zpt\ distribution on boson rapidity~\cite{coordinates}. 

\begin{figure}[htbp]
    \includegraphics[width=0.45\textwidth]{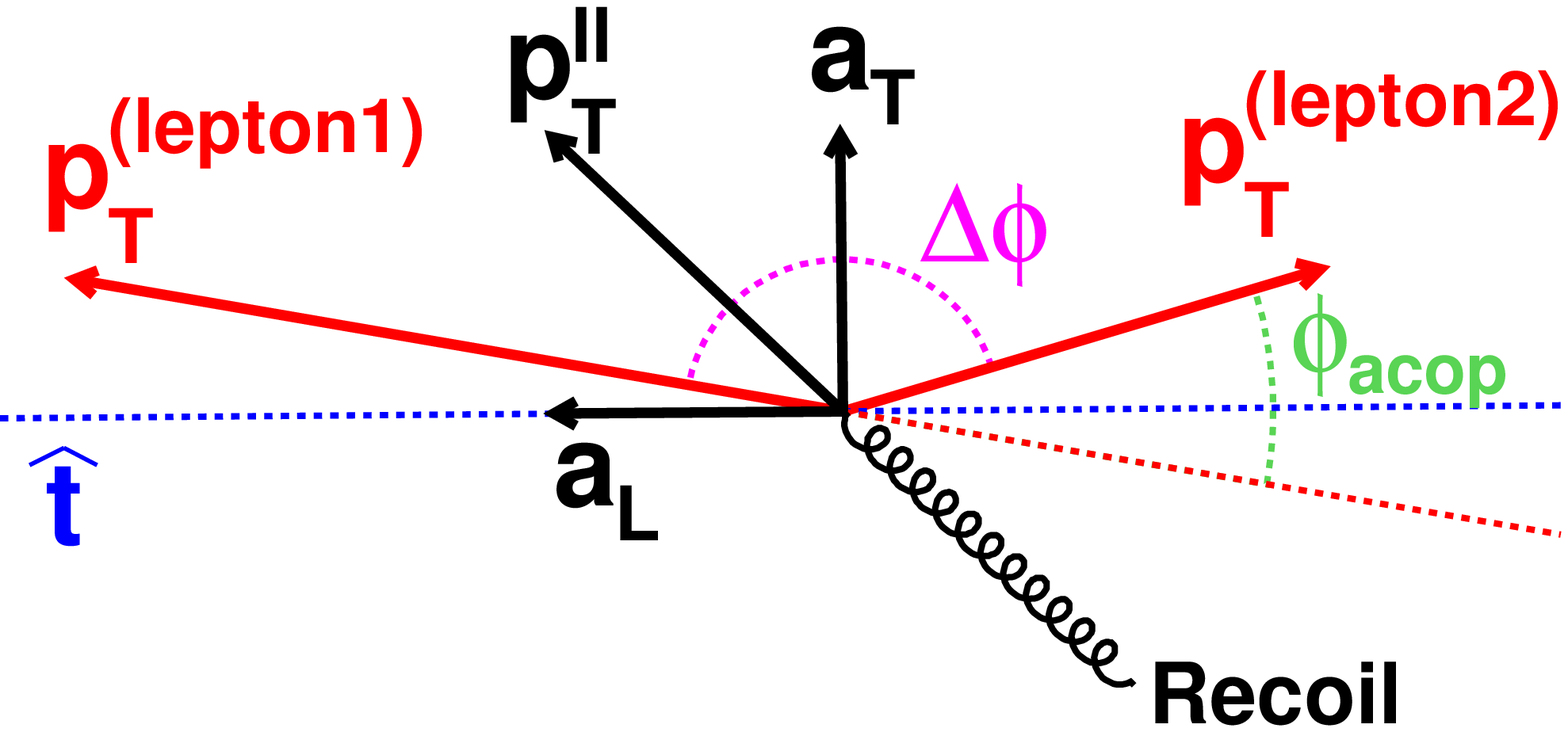}\\
    \caption{Illustration of the variables defined in the
      text and used to analyse the dilepton transverse momentum.}
    \label{Figure:at}
  \end{figure}

In the region of low  \zpt, the precision of the most recent measurements
at the Tevatron~\cite{DzeroRunIIa,DzeroDimuZpT}
was dominated by uncertainties in correcting for  experimental resolution and efficiency.
Furthermore, the choice of bin widths was restricted by experimental
resolution rather than event statistics.
The variable \at, which corresponds to the component
of \zpt\ that is transverse to the dilepton thrust axis, $\hat{t}$, has been
proposed as an alternative analysing variable that
allows us to study the issues discussed above, but is less susceptible than the \zpt\ to detector effects~\cite{aT_paper}.
Figure~\ref{Figure:at} illustrates this and other relevant variables
defined below.
The \at\ distribution was subsequently calculated to next-to-leading-log (NLL) accuracy using resummation techniques~\cite{Banfi_at}.
Additional analysing variables with even better experimental
resolution have recently been proposed and studied~\cite{aToM_paper}.
The optimal variable was found to be  \phistar, which is defined as:
$$\phistar = \tan\left(\phiaco/2\right)\sin(\theta^*_{\eta}),$$
where \phiaco\ is the acoplanarity angle, given by:
$\phiaco = \pi - \Delta\phi^{\ell\ell}$, 
and $\Delta\phi^{\ell\ell}$ is the difference in azimuthal angle, $\phi$, between the two lepton
candidates.
The variable $\theta^*_{\eta}$ is a measure of the scattering angle of the leptons
with respect to the proton beam direction in the rest frame of the dilepton system.
It is defined~\cite{aToM_paper} by:
$\cos(\theta^{*}_{\eta})=\tanh\left[\left(\eta^--\eta^+\right)/2\right]$, 
where $\eta^-$ and $\eta^+$ are the pseudorapidities~\cite{coordinates} 
 of the negatively
and positively charged lepton, respectively.

The variable \phistar\ is highly correlated
with the quantity $\at/m_{\ell\ell}$, where $m_{\ell\ell}$ is the dilepton invariant mass.
Since \phiaco\ and $\theta^{*}_{\eta}$ depend exclusively on the directions of the two
leptons, which are measured with a precision of a milliradian or better, \phistar\ is experimentally very well measured compared
to any quantities that rely on the momenta  of the leptons.

We present a measurement of the normalized \phistar\ distribution,
\oneoversigmaphistar, in bins of $|y|$,
using \lumi~fb$^{-1}$ of $p\bar{p}$
  collisions collected by the D0 detector at the Fermilab Tevatron.
The \phistar\ distributions are measured in both dielectron and dimuon
  events and are corrected for  experimental resolution and efficiency.
We correct back to
the level of observable, generator-level particles; that is, we apply kinematic
selection criteria at the particle level
that match those applied in the selection of candidate events in the data~\cite{FSR_paper}.
Particle level electrons are defined as the four-vector sum of any electrons and photons
within a cone of $\Delta{\cal R} = \sqrt{(\Delta\eta)^2 +
  (\Delta\phi)^2}$~$<$~0.2 around an electron, where 
$\Delta\eta$~$(\Delta\phi)$ is the distance in $\eta$~$(\phi)$ from the
particle level  electron;
this mimics the measurement of electron energy in the calorimeter.
Particle level muons are defined after QED final state radiation;
this mimics the measurement of muon momentum in the tracking detector.
The  kinematic selection criteria  are:
electrons must satisfy \pt~$>$~20~GeV and $|\eta|$~$<$~1.1 or 1.5~$<$~$|\eta|$~$<$~3;
muons must satisfy $p_T$~$>$~15 GeV and
$|\eta|$~$<$~2;
$m_{\ell\ell}$  must fall within the range 70--110~GeV.
%and the appropriate $|y|$ range,

The corrected data are compared to predictions from the Monte Carlo (MC)
program \resbos~\cite{ResBos} with the above kinematic selection criteria
applied at the particle level.
\resbos\ generates $Z/\gamma^*$ boson
events with initial state QCD corrections to next-to-leading order
 (NLO) and NLL accuracy together with: a non-perturbative form factor, whose width
% in high energy hadron-hadron collisions
 is controlled primarily by the parameter  
$g_2$ (with default value $[0.68^{+0.02}_{-0.01}]$~GeV$^2$)\cite{BLNY2003}; an additional next-to-NLO
(NNLO) K-factor~\cite{NNLO_KF};
 CTEQ6.6 NLO parton distribution functions (PDFs)~\cite{CTEQ66};
and QED radiative corrections from \photos~\cite{photos}.
The QCD factorization and
renormalization scales are set event by event to the mass of the
$Z/\gamma^*$ boson propagator.

The D0 detector~\cite{DetectorNIM} consists of: 
silicon microstrip  and central fiber tracking detectors,
located within a 2~T superconducting solenoid;
a liquid-argon/uranium sampling calorimeter; and an outer muon system
consisting of tracking and scintillation detectors 
located before and after 1.8~T toroids.
Candidate dielectron events are required to satisfy a
trigger based on the identification of a single electron and to contain two clusters
reconstructed in the calorimeter with a transverse and longitudinal
shower profile consistent with that expected of an electron.
The calorimeter is housed in three separate cryostats; this has the effect that
 electron identification is degraded in the region \mbox{$1.1<|\eta|<1.5$}.
Candidate dimuon events are required to satisfy a
trigger based on the identification of a single muon and to contain
two muons reconstructed either in the outer
muon system, or as an energy deposit consistent with the passage of a
minimum-ionizing particle in the calorimeter.
In order to ensure an accurate measurement of the lepton directions at
the point of production, the two lepton candidates are required to be matched to
a pair of oppositely charged particle tracks reconstructed in the central tracking detectors.
%the two matched tracks are required to be of opposite charge.
Candidate leptons resulting from misidentified hadrons or produced by the decay of hadrons 
are suppressed by requiring that they be isolated from other particles
in the event and, in the case of electrons with $|\eta|<1.1$, by requiring
the energy measured in the calorimeter and the momentum measured in
the central tracking detectors to be consistent.
Contamination from cosmic ray muons is strongly suppressed by a requirement that
the muons originate from the  $p\bar{p}$
  collision point and by rejecting events
in which the two muon candidates 
are back to back in $\eta$.
In total,  455k dielectron events and 511k dimuon events are selected.

The corrections to the observed \phistar\ distribution for experimental resolution and efficiency are evaluated
using $Z/\gamma^*$ boson MC events that are generated with \pythia~\cite{PYTHIA} and passed through a \geant-based~\cite{GEANT}  simulation of the detector.
These fully simulated MC events are re-weighted at the generator level in two dimensions (\zpt\ and $|y|$) to match the predictions
of \resbos.
In addition, adjustments are made to improve the accuracy of the
following aspects of the detector simulation: electron energy and muon $p_T$ scale and resolution;
track $\phi$ and $\eta$ resolutions;
trigger efficiencies; and 
%(which are not directly simulated)
relevant offline reconstruction and selection
efficiencies.
Variations in the above adjustments to the underlying physics and the
detector simulation are included in the assessment of the systematic
uncertainties on the correction factors.

The systematic uncertainties due to electron energy and muon $p_T$ scale
and resolution are small, and arise only due to the kinematic
requirements in the event selection.
The measured \phistar\ distribution is, however, susceptible to
modulations in $\phi$ of the
lepton identification and trigger efficiencies, which result, e.g.,  from
detector module boundaries in the  calorimeter and muon systems.
Particular care has been taken (a)~in the choice of lepton identification
criteria in order to minimize such modulations and (b)~to ensure that such modulations are
well simulated in the MC.
For example, the requirements imposed on shower profile are much
looser than those usually employed in electron identification within D0,
because tight requirements are particularly inefficient in the
regions close to module boundaries in the calorimeter.
Similarly, the inclusion of muon candidates identified in the
calorimeter reduces the effect of gaps between modules in the outer
muon system.
Accurate modelling of the angular resolution of the central tracking
detectors is another crucial aspect of this analysis. 
The resolution in $\phi$ and $\eta$ is measured in the data using
cosmic ray muons that traverse the detector, since these should
produce events containing two tracks that are exactly back to back
except for the effect of detector resolution.  
%Modelling of the inefficiencies in so called $\phi$-gaps is particularly important in this measurement,
%since regions of low efficiency which are back-to-back in $\phi$ 
%cause the efficiency to modulate as a function of \phistar. 
%The simulation is verified to accurately describe the inefficiencies in the relevant 
%$\phi$-gaps of the central calorimeter, and outer muon system.
%A final generator level re-weighting corrects the phistar\ distribution to
%that measured in a previous iteration of the analysis.

Backgrounds from \ztt, \mbox{$W\rightarrow \ell\nu$}~(+jets), and $WW\rightarrow \ell\nu\ell\nu$
are simulated using \pythia.
Background from top quark pair events is simulated with
\alpgen~\cite{ALPGEN}, with \pythia\ used for parton showering. 
Background from multijet events is estimated from data.
The total fraction of background events is 0.26\% for the dielectron channel, and 0.38\% 
for the dimuon channel.

%After verifying that the 
%fully simulated MC event sample describes the data with sufficient
%accuracy, it can be used to correct the data back to the particle level.
Since the experimental resolution in \phistar\ is narrower than the
chosen bin widths, the
fractions of accepted events that fall within the same bin in  \phistar\ at the
particle level and reconstructed detector level in the MC are high, having
typical (lowest) values of around 98\% (92\%). 
Therefore, simple bin-by-bin corrections of the \phistar\ distribution
are sufficient.
%The systematic uncertainties in the correction factors arising from
%variations in the assumed underlying  \phistar\ distribution are found
%to be small.
%Other systematic uncertainties on the measured distributions arise from 
%the modelling of the detector response in the simulation.
%The following are varied within their uncertainties:
%electron energy and muon $p_T$ scales and resolutions;
%dependence of trigger and offline identification efficiencies
%on $\eta$ and on the proximity to detector module boundaries in
%$\phi$.
In almost all \phistar\ bins the total systematic uncertainty is
substantially smaller than the statistical uncertainty.
%FIXME, discuss the physics input biases. FIXME also need to describe the \phistar\ dependent K-factor earlier.
%FIXME: quantitative comparison of the average size of stat and syst uncertainties.

\begin{figure*}[hbtp]
\begin{minipage}[b]{0.97\textwidth}
\centering
\includegraphics[width=0.996\textwidth]{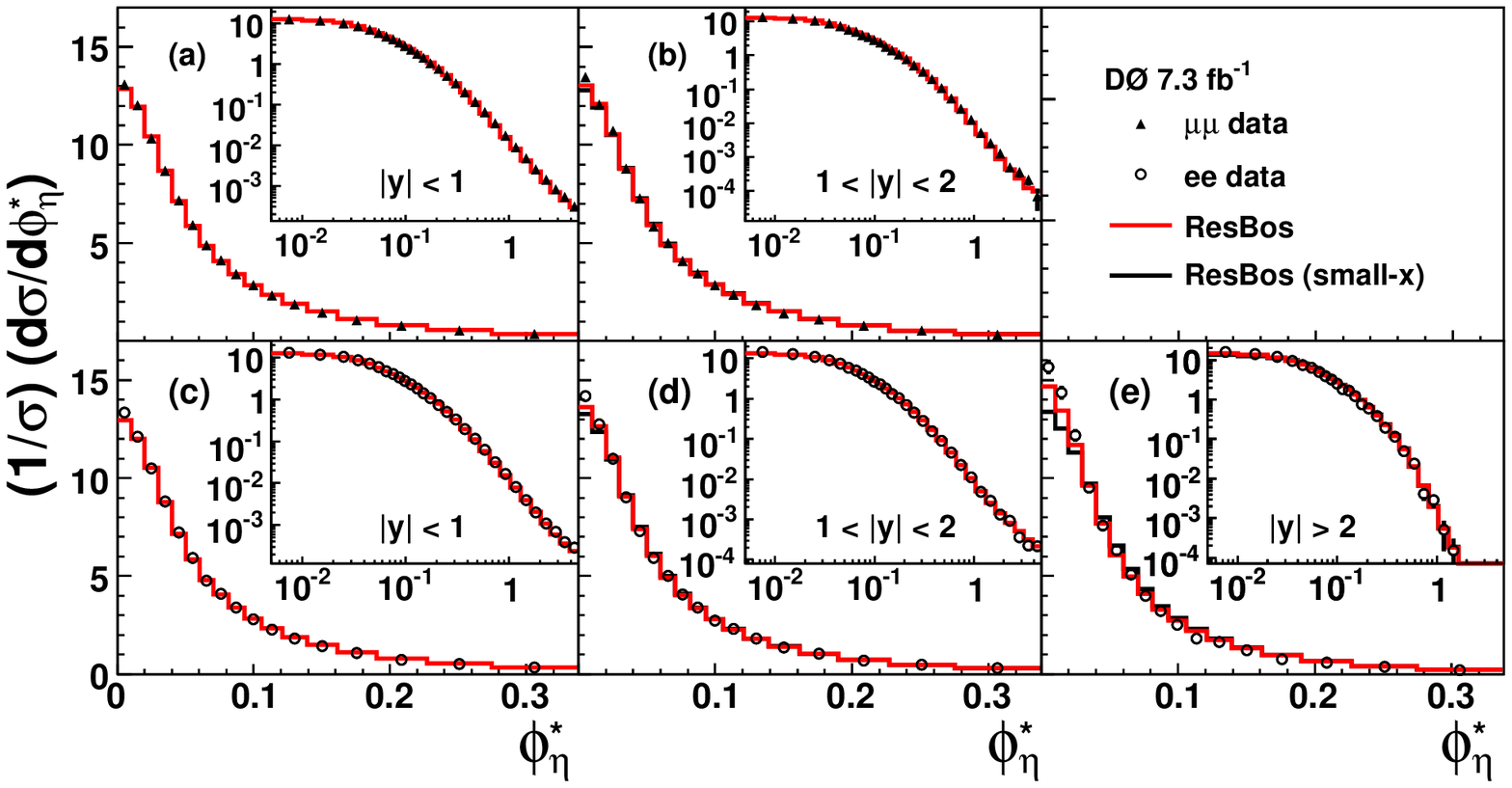}
\caption{(color online) Corrected distributions of
  \oneoversigmaphistar\ for dimuon events with (a)~$|y| < 1$ and
  (b)~$1 < |y| < 2$; and  dielectron events with (c)~$|y| < 1$, 
  (d)~$1 < |y| < 2$  and (e)~$|y| > 2$.
  The larger plots show the
  restricted range $0 < \phistar < 0.34$ and the insets show the 
  full range of \phistar.
%The error bars on the data points represent statistical and systematic
%uncertainties combined in quadrature.
The predictions from  \resbos\ are shown as the red histogram
  and from \resbos\ with small-$x$ broadening as the black histogram
  [which is visible principally in~(e)].
}
\label{fig:figure1}
\end{minipage}
\end{figure*}

\begin{figure*}[hbtp]
\begin{minipage}[b]{0.97\textwidth}
\centering
\includegraphics[width=0.996\textwidth]{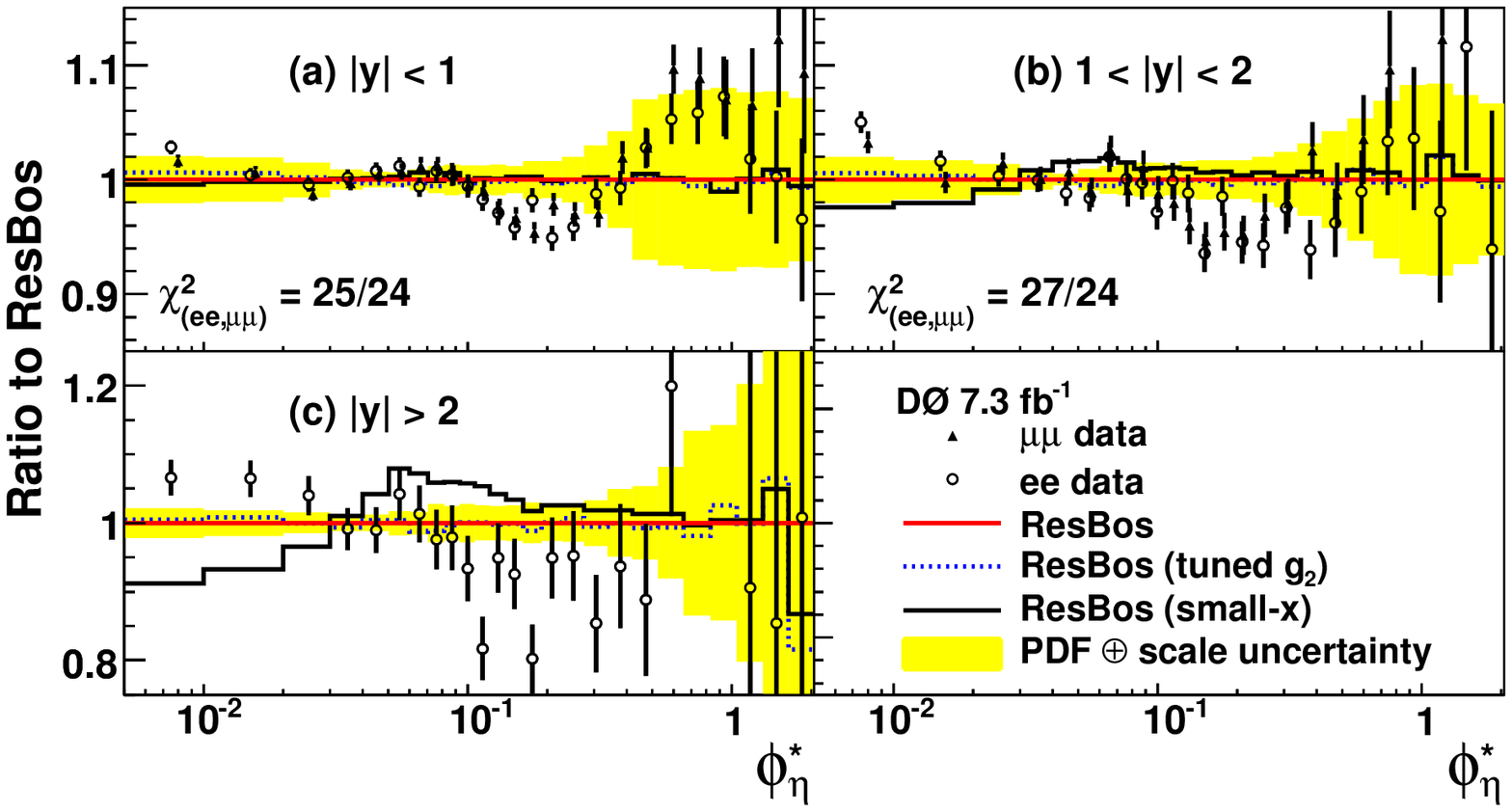}
\caption{(color online) Ratio of the corrected distributions of
  \oneoversigmaphistar\ to 
   \resbos\  for: (a)~$|y| < 1$,
  (b)~$1$~$<$~$|y|$~$<$~2 and (c)~$|y| > 2$.
Statistical and systematic
uncertainties are combined in quadrature.
In~(a) and~(b) a $\chi^2$ for the comparison of the dielectron and
  dimuon data, $\chi^2_{(ee/\mu\mu)}$, is calculated assuming
  uncorrelated uncertainties.
The yellow band around the \resbos\ prediction represents the
quadrature sum of 
uncertainty due to PDFs (evaluated using the CTEQ6.6
NLO error PDFs~\protect\cite{CTEQ66}) and the uncertainty due to the
QCD scale (evaluated by varying the factorization and
renormalization scales simultaneously by a factor of two).
Also shown are the changes to the \resbos\ predictions when 
$g_2$ is set to 0.66 (dotted blue line) and when the small-$x$ broadening option
is enabled (solid black line). 
}
\label{fig:figure2}
\end{minipage}
\end{figure*}

Figure~\ref{fig:figure1}
shows the corrected particle level \phistar\ distributions together with predictions from
\resbos.
%In all cases, the distributions are normalized to unit area. %(within the presented \phistar\ range).
Figure~\ref{fig:figure2} shows the ratio
of the corrected \phistar\ distributions to the \resbos\ predictions in
both the dielectron and  dimuon channels.
The general shape of the distributions is broadly
described by \resbos\ over the full range in \phistar.
However, the small statistical uncertainties resulting from the large
dilepton data sets, combined with the fine binning and small systematic
uncertainties resulting from the use of \phistar\ as the analysing
variable, reveal differences between the data and  \resbos.
Since the particle level definitions for
electrons and muons to which the data are
corrected are slightly different,
Fig.~\ref{fig:figure2} represents the most appropriate way
to demonstrate the  consistency of  the dielectron and dimuon data.
%The corrected dielectron and dimuon data are consistent with one
%another and deviate in a consistent fashion from the MC predictions.
Given that the experimental acceptance corrections are very different
in the two channels, this consistency represents a powerful cross check of the
corrected distributions.

%In order to allow \resbos\ the best possible chance to describe the data,
%the $g_2$ parameter is allowed to float.
The results of fits for the  value of $g_2$, separately 
in each $|y|$ bin and channel, are shown in Table~\ref{Table:g2fits}.
It can be seen that the fitted values of \gtwo\ show a monotonic decrease 
with increasing $|y|$ for both channels.
That is, the width of the \phistar\ distribution becomes  narrower
with increasing $|y|$ faster in the data than is predicted by \resbos.
This is the opposite of the behavior expected from the
small-$x$ broadening hypothesis~\cite{Nadolsky-smallx-2001,BergeSmallx-2005}.
Figure~\ref{fig:figure2} confirms that 
the prediction from \resbos\ with small-$x$ broadening is in  poor 
agreement with data.
%, especially in regionterms of the $y$ dependence.
It can also be seen that choosing the $g_2$ value (0.66~GeV$^2$)
that best describes the average behavior of the data over all  $|y|$ bins and channels
has  little effect on the level of agreement with data.
  \begin{table}[htb]
    \centering
    \begin{tabular}{c|c|c|c}
      \hline \hline
      Channel & $|y| < 1$ & $1 < |y| < 2$ & $|y| > 2$ \\
      \hline
      $ee$      &  0.644 $\pm$ 0.013 & 0.619 $\pm$ 0.017 & 0.550 $\pm$ 0.048 \\
      $\mu\mu$  &  0.670 $\pm$ 0.012 & 0.645 $\pm$ 0.019 & -- \\
      \hline \hline
    \end{tabular}
    \caption{Value of $g_2$ (GeV$^2$) that best describes the corrected data
      in each $|y|$ bin and channel.}
    \label{Table:g2fits}
  \end{table}

A previous measurement~\cite{DzeroDimuZpT} showed that, for
central rapidities,
\resbos\ underestimates the number of $Z/\gamma^*$ bosons at high
\zpt\ by about 10\%.
This is consistent with the deviations seen at high values of
\phistar\ in Fig.~\ref{fig:figure2}~(a). 

In summary, using \lumi~\invfb\ of  $p\bar{p}$
  collisions collected by the D0 detector at the Fermilab Tevatron, 
we have studied with unprecedented precision the \zpt\ distribution of $Z/\gamma^*$ bosons
in dielectron and dimuon final states.
In bins of boson rapidity, the normalised cross section 
is measured as a function of the variable \phistar. 
%which is sensitive to the same physics as the \zpt\ distribution,
%but less susceptible to detector resolution and efficiency effects.
Predictions from  {\sc ResBos} are unable to describe the detailed
shape of the corrected data,
and a prediction that includes the effect of small-$x$ broadening
is  strongly disfavored.

Tables of corrected \oneoversigmaphistar\ distributions
for each $|y|$ bin and channel are provided~\cite{Supplementary}.

% acknowledgement.tex                             17 May 2010
%
We thank the staffs at Fermilab and collaborating institutions,
and acknowledge support from the
DOE and NSF (USA);
CEA and CNRS/IN2P3 (France);
FASI, Rosatom and RFBR (Russia);
CNPq, FAPERJ, FAPESP and FUNDUNESP (Brazil);
DAE and DST (India);
Colciencias (Colombia);
CONACyT (Mexico);
KRF and KOSEF (Korea);
CONICET and UBACyT (Argentina);
FOM (The Netherlands);
STFC and the Royal Society (United Kingdom);
MSMT and GACR (Czech Republic);
CRC Program and NSERC (Canada);
BMBF and DFG (Germany);
SFI (Ireland);
The Swedish Research Council (Sweden);
and
CAS and CNSF (China).
%
   % input acknowledgement

\clearpage
%\newpage
\begin{center}
{\large\bf Supplementary Material}
\end{center}

Tables~\ref{Table:results_diem_y1}--\ref{Table:results_dimu_y2} show
the values of the corrected \oneoversigmaphistar\ distributions
for each $|y|$ bin and channel.
The first uncertainty is statistical and the second is systematic. 
The data are corrected back to the particle level, corresponding to
the kinematic cuts and definitions of particle level leptons as
described in the text.
The integral of
\oneoversigmaphistar\ for \phistar\ in the range 0--4.749 is
normalised to unity separately for
each $|y|$ bin and channel, with the exception of the dielectron
channel for $|y|>2$ in which the integral is normalised in the
\phistar\ range 0--2.049 .

%  \begin{table}[hbpt]
%     \caption{Table of results for the dielectron channel and inclusive rapidity.}
%   \centering
%   \begin{tabular}{  c c c  }
%    \hline \hline
%   bin & range & $1/\sigma$~$d\sigma/d\phistarEta$ \\
%   \hline
%   \input{results_table_diem_y0.tex}
%   \hline \hline
%   \end{tabular}
%   \label{Table:results_diem_y0}
%   \end{table}

  \begin{table}[hbpt]
       \caption{The corrected ($1/\sigma$)~$\times$~($d\sigma/d\phistarEta$) distribution for the dielectron channel and $|y| < 1$.}
   \centering
   \begin{tabular}{  c c c  }
      \hline \hline
   bin & range & ($1/\sigma$)~$\times$~($d\sigma/d\phistarEta$) \\
   \hline
   1 & 0.000-0.010 & 13.242 $\pm$ 0.065 $\pm$ 0.020\\
2 & 0.010-0.020 & 12.006 $\pm$ 0.062 $\pm$ 0.012\\
3 & 0.020-0.030 & 10.429 $\pm$ 0.058 $\pm$ 0.012\\
4 & 0.030-0.040 & 8.756 $\pm$ 0.053 $\pm$ 0.008\\
5 & 0.040-0.050 & 7.183 $\pm$ 0.048 $\pm$ 0.007\\
6 & 0.050-0.060 & 5.911 $\pm$ 0.043 $\pm$ 0.005\\
7 & 0.060-0.071 & 4.762 $\pm$ 0.039 $\pm$ 0.005\\
8 & 0.071-0.081 & 4.070 $\pm$ 0.035 $\pm$ 0.003\\
9 & 0.081-0.093 & 3.387 $\pm$ 0.031 $\pm$ 0.005\\
10 & 0.093-0.106 & 2.806 $\pm$ 0.026 $\pm$ 0.004\\
11 & 0.106-0.121 & 2.279 $\pm$ 0.022 $\pm$ 0.003\\
12 & 0.121-0.139 & 1.830 $\pm$ 0.018 $\pm$ 0.003\\
13 & 0.139-0.162 & 1.414 $\pm$ 0.014 $\pm$ 0.003\\
14 & 0.162-0.190 & 1.084 $\pm$ 0.011 $\pm$ 0.002\\
15 & 0.190-0.227 & 0.750 $\pm$ 0.008 $\pm$ 0.002\\
16 & 0.227-0.275 & 0.513 $\pm$ 0.006 $\pm$ 0.001\\
17 & 0.275-0.337 & 0.333 $\pm$ 0.004 $\pm$ 0.001\\
18 & 0.337-0.418 & 0.197 $\pm$ 0.003 $\pm$ 0.000\\
19 & 0.418-0.523 & 0.115 $\pm$ 0.002 $\pm$ 0.000\\
\hline
bin & range & ($1/\sigma$)~$\times$~($d\sigma/d\phistar$) ($\times 1000$) \\
\hline
20 & 0.523-0.657 & 61.731 $\pm$ 1.238 $\pm$ 0.090\\
21 & 0.657-0.827 & 32.115 $\pm$ 0.798 $\pm$ 0.080\\
22 & 0.827-1.041 & 16.496 $\pm$ 0.509 $\pm$ 0.071\\
23 & 1.041-1.309 & 7.960 $\pm$ 0.324 $\pm$ 0.171\\
24 & 1.309-1.640 & 3.882 $\pm$ 0.203 $\pm$ 0.087\\
25 & 1.640-2.049 & 2.006 $\pm$ 0.133 $\pm$ 0.056\\
26 & 2.049-2.547 & 1.068 $\pm$ 0.090 $\pm$ 0.033\\
27 & 2.547-3.151 & 0.702 $\pm$ 0.067 $\pm$ 0.028\\
28 & 3.151-3.878 & 0.389 $\pm$ 0.045 $\pm$ 0.015\\
29 & 3.878-4.749 & 0.284 $\pm$ 0.036 $\pm$ 0.013\\

   \hline \hline
   \end{tabular}
   \label{Table:results_diem_y1}
   \end{table}

   \begin{table}[hbpt]
       \caption{The corrected ($1/\sigma$)~$\times$~($d\sigma/d\phistarEta$) distribution for the dielectron channel and $1 < |y| < 2$.}
   \centering
   \begin{tabular}{  c c c  }
      \hline \hline
   bin & range & ($1/\sigma$)~$\times$~($d\sigma/d\phistarEta$) \\
   \hline
   1 & 0.000-0.010 & 14.235 $\pm$ 0.104 $\pm$ 0.053\\
2 & 0.010-0.020 & 12.782 $\pm$ 0.099 $\pm$ 0.046\\
3 & 0.020-0.030 & 11.035 $\pm$ 0.092 $\pm$ 0.030\\
4 & 0.030-0.040 & 9.023 $\pm$ 0.083 $\pm$ 0.019\\
5 & 0.040-0.050 & 7.268 $\pm$ 0.074 $\pm$ 0.011\\
6 & 0.050-0.060 & 5.911 $\pm$ 0.067 $\pm$ 0.007\\
7 & 0.060-0.071 & 4.988 $\pm$ 0.061 $\pm$ 0.006\\
8 & 0.071-0.081 & 4.029 $\pm$ 0.053 $\pm$ 0.007\\
9 & 0.081-0.093 & 3.346 $\pm$ 0.047 $\pm$ 0.006\\
10 & 0.093-0.106 & 2.696 $\pm$ 0.040 $\pm$ 0.008\\
11 & 0.106-0.121 & 2.267 $\pm$ 0.034 $\pm$ 0.010\\
12 & 0.121-0.139 & 1.780 $\pm$ 0.027 $\pm$ 0.010\\
13 & 0.139-0.162 & 1.308 $\pm$ 0.021 $\pm$ 0.007\\
14 & 0.162-0.190 & 1.015 $\pm$ 0.016 $\pm$ 0.005\\
15 & 0.190-0.227 & 0.683 $\pm$ 0.012 $\pm$ 0.005\\
16 & 0.227-0.275 & 0.444 $\pm$ 0.008 $\pm$ 0.003\\
17 & 0.275-0.337 & 0.282 $\pm$ 0.006 $\pm$ 0.003\\
18 & 0.337-0.418 & 0.157 $\pm$ 0.004 $\pm$ 0.002\\
19 & 0.418-0.523 & 0.089 $\pm$ 0.003 $\pm$ 0.001\\
\hline
bin & range & ($1/\sigma$)~$\times$~($d\sigma/d\phistar$) ($\times 1000$) \\
\hline
20 & 0.523-0.657 & 45.296 $\pm$ 1.581 $\pm$ 0.371\\
21 & 0.657-0.827 & 22.931 $\pm$ 1.019 $\pm$ 0.113\\
22 & 0.827-1.041 & 10.886 $\pm$ 0.636 $\pm$ 0.092\\
23 & 1.041-1.309 & 4.909 $\pm$ 0.378 $\pm$ 0.114\\
24 & 1.309-1.640 & 2.848 $\pm$ 0.262 $\pm$ 0.073\\
25 & 1.640-2.049 & 1.330 $\pm$ 0.161 $\pm$ 0.045\\
26 & 2.049-2.547 & 0.921 $\pm$ 0.120 $\pm$ 0.044\\
27 & 2.547-3.151 & 0.363 $\pm$ 0.069 $\pm$ 0.019\\
28 & 3.151-3.878 & 0.226 $\pm$ 0.051 $\pm$ 0.010\\
29 & 3.878-4.749 & 0.214 $\pm$ 0.048 $\pm$ 0.013\\

   \hline \hline
   \end{tabular}
   \label{Table:results_diem_y2}
   \end{table}

\begin{table}[hbpt]
       \caption{The corrected ($1/\sigma$)~$\times$~($d\sigma/d\phistarEta$) distribution for the dielectron channel and $|y| > 2$.}
   \centering
   \begin{tabular}{  c c c  }
      \hline \hline
   bin & range & ($1/\sigma$)~$\times$~($d\sigma/d\phistarEta$) \\
   \hline
   1 & 0.000-0.010 & 15.625 $\pm$ 0.361 $\pm$ 0.031\\
2 & 0.010-0.020 & 14.288 $\pm$ 0.344 $\pm$ 0.023\\
3 & 0.020-0.030 & 12.130 $\pm$ 0.319 $\pm$ 0.030\\
4 & 0.030-0.040 & 9.514 $\pm$ 0.281 $\pm$ 0.027\\
5 & 0.040-0.050 & 7.572 $\pm$ 0.250 $\pm$ 0.009\\
6 & 0.050-0.060 & 6.311 $\pm$ 0.226 $\pm$ 0.016\\
7 & 0.060-0.071 & 5.052 $\pm$ 0.202 $\pm$ 0.013\\
8 & 0.071-0.081 & 3.991 $\pm$ 0.175 $\pm$ 0.008\\
9 & 0.081-0.093 & 3.206 $\pm$ 0.152 $\pm$ 0.008\\
10 & 0.093-0.106 & 2.533 $\pm$ 0.126 $\pm$ 0.006\\
11 & 0.106-0.121 & 1.796 $\pm$ 0.099 $\pm$ 0.006\\
12 & 0.121-0.139 & 1.658 $\pm$ 0.087 $\pm$ 0.006\\
13 & 0.139-0.162 & 1.223 $\pm$ 0.067 $\pm$ 0.002\\
14 & 0.162-0.190 & 0.767 $\pm$ 0.047 $\pm$ 0.005\\
15 & 0.190-0.227 & 0.605 $\pm$ 0.037 $\pm$ 0.003\\
16 & 0.227-0.275 & 0.378 $\pm$ 0.025 $\pm$ 0.003\\
17 & 0.275-0.337 & 0.195 $\pm$ 0.016 $\pm$ 0.002\\
18 & 0.337-0.418 & 0.114 $\pm$ 0.011 $\pm$ 0.001\\
19 & 0.418-0.523 & 0.050 $\pm$ 0.006 $\pm$ 0.001\\
\hline
bin & range & ($1/\sigma$)~$\times$~($d\sigma/d\phistar$) ($\times 1000$) \\
\hline
20 & 0.523-0.657 & 24.457 $\pm$ 3.781 $\pm$ 0.616\\
21 & 0.657-0.827 & 4.145 $\pm$ 1.382 $\pm$ 0.122\\
22 & 0.827-1.041 & 2.906 $\pm$ 1.099 $\pm$ 0.148\\
23 & 1.041-1.309 & 0.503 $\pm$ 0.356 $\pm$ 0.024\\
24 & 1.309-1.640 & 0.154 $\pm$ 0.154 $\pm$ 0.014\\
25 & 1.640-2.049 & 0.101 $\pm$ 0.101 $\pm$ 0.009\\

   \hline \hline
   \end{tabular}
   \label{Table:results_diem_y3}
   \end{table}

%  \begin{table}[hbpt]
%     \caption{The corrected ($1/\sigma$)~$\times$~($d\sigma/d\phistarEta$) distribution for the dimuon channel and inclusive rapidity.}
%   \centering
%   \begin{tabular}{  c c c  }
%      \hline \hline
%   bin & range & ($1/\sigma$)~$\times$~($d\sigma/d\phistarEta$) \\
%   \hline
%   \input{results_table_dimu_y0.tex}
%   \hline \hline
%   \end{tabular}
%   \label{Table:results_dimu_y0}
%   \end{table}

  \begin{table}[hbpt]
       \caption{The corrected ($1/\sigma$)~$\times$~($d\sigma/d\phistarEta$) distribution for the dimuon channel and $|y| < 1$.}
   \centering
   \begin{tabular}{  c c c  }
      \hline \hline
   bin & range & ($1/\sigma$)~$\times$~($d\sigma/d\phistarEta$) \\
   \hline
   1 & 0.000-0.010 & 12.992 $\pm$ 0.058 $\pm$ 0.027\\
2 & 0.010-0.020 & 11.958 $\pm$ 0.055 $\pm$ 0.022\\
3 & 0.020-0.030 & 10.263 $\pm$ 0.051 $\pm$ 0.015\\
4 & 0.030-0.040 & 8.620 $\pm$ 0.047 $\pm$ 0.013\\
5 & 0.040-0.050 & 7.115 $\pm$ 0.043 $\pm$ 0.010\\
6 & 0.050-0.060 & 5.863 $\pm$ 0.039 $\pm$ 0.007\\
7 & 0.060-0.071 & 4.868 $\pm$ 0.035 $\pm$ 0.006\\
8 & 0.071-0.081 & 4.084 $\pm$ 0.031 $\pm$ 0.006\\
9 & 0.081-0.093 & 3.400 $\pm$ 0.027 $\pm$ 0.005\\
10 & 0.093-0.106 & 2.831 $\pm$ 0.024 $\pm$ 0.004\\
11 & 0.106-0.121 & 2.320 $\pm$ 0.020 $\pm$ 0.003\\
12 & 0.121-0.139 & 1.850 $\pm$ 0.016 $\pm$ 0.003\\
13 & 0.139-0.162 & 1.439 $\pm$ 0.013 $\pm$ 0.003\\
14 & 0.162-0.190 & 1.061 $\pm$ 0.010 $\pm$ 0.002\\
15 & 0.190-0.227 & 0.779 $\pm$ 0.007 $\pm$ 0.002\\
16 & 0.227-0.275 & 0.526 $\pm$ 0.005 $\pm$ 0.002\\
17 & 0.275-0.337 & 0.331 $\pm$ 0.004 $\pm$ 0.001\\
18 & 0.337-0.418 & 0.207 $\pm$ 0.003 $\pm$ 0.001\\
19 & 0.418-0.523 & 0.117 $\pm$ 0.002 $\pm$ 0.001\\
\hline
bin & range & ($1/\sigma$)~$\times$~($d\sigma/d\phistar$) ($\times 1000$) \\
\hline
20 & 0.523-0.657 & 66.026 $\pm$ 1.160 $\pm$ 0.479\\
21 & 0.657-0.827 & 34.188 $\pm$ 0.753 $\pm$ 0.304\\
22 & 0.827-1.041 & 17.056 $\pm$ 0.480 $\pm$ 0.256\\
23 & 1.041-1.309 & 8.753 $\pm$ 0.315 $\pm$ 0.254\\
24 & 1.309-1.640 & 4.774 $\pm$ 0.214 $\pm$ 0.130\\
25 & 1.640-2.049 & 2.489 $\pm$ 0.139 $\pm$ 0.073\\
26 & 2.049-2.547 & 1.394 $\pm$ 0.096 $\pm$ 0.044\\
27 & 2.547-3.151 & 0.802 $\pm$ 0.066 $\pm$ 0.025\\
28 & 3.151-3.878 & 0.536 $\pm$ 0.051 $\pm$ 0.025\\
29 & 3.878-4.749 & 0.307 $\pm$ 0.034 $\pm$ 0.012\\

   \hline \hline
   \end{tabular}
   \label{Table:results_dimu_y1}
   \end{table}

   \begin{table}[hbpt]
       \caption{The corrected ($1/\sigma$)~$\times$~($d\sigma/d\phistarEta$) distribution for the dimuon channel and $1 < |y| < 2$.}
   \centering
   \begin{tabular}{  c c c  }
      \hline \hline
   bin & range & ($1/\sigma$)~$\times$~($d\sigma/d\phistarEta$) \\
   \hline
   1 & 0.000-0.010 & 13.404 $\pm$ 0.105 $\pm$ 0.051\\
2 & 0.010-0.020 & 12.008 $\pm$ 0.100 $\pm$ 0.026\\
3 & 0.020-0.030 & 10.647 $\pm$ 0.094 $\pm$ 0.016\\
4 & 0.030-0.040 & 8.755 $\pm$ 0.086 $\pm$ 0.013\\
5 & 0.040-0.050 & 7.230 $\pm$ 0.077 $\pm$ 0.016\\
6 & 0.050-0.060 & 5.804 $\pm$ 0.069 $\pm$ 0.019\\
7 & 0.060-0.071 & 4.972 $\pm$ 0.063 $\pm$ 0.011\\
8 & 0.071-0.081 & 4.045 $\pm$ 0.056 $\pm$ 0.008\\
9 & 0.081-0.093 & 3.441 $\pm$ 0.050 $\pm$ 0.008\\
10 & 0.093-0.106 & 2.820 $\pm$ 0.043 $\pm$ 0.009\\
11 & 0.106-0.121 & 2.330 $\pm$ 0.036 $\pm$ 0.006\\
12 & 0.121-0.139 & 1.824 $\pm$ 0.029 $\pm$ 0.004\\
13 & 0.139-0.162 & 1.414 $\pm$ 0.023 $\pm$ 0.003\\
14 & 0.162-0.190 & 1.066 $\pm$ 0.018 $\pm$ 0.002\\
15 & 0.190-0.227 & 0.756 $\pm$ 0.013 $\pm$ 0.001\\
16 & 0.227-0.275 & 0.514 $\pm$ 0.009 $\pm$ 0.002\\
17 & 0.275-0.337 & 0.326 $\pm$ 0.007 $\pm$ 0.002\\
18 & 0.337-0.418 & 0.200 $\pm$ 0.005 $\pm$ 0.001\\
19 & 0.418-0.523 & 0.107 $\pm$ 0.003 $\pm$ 0.000\\
\hline
bin & range & ($1/\sigma$)~$\times$~($d\sigma/d\phistar$) ($\times 1000$) \\
\hline
20 & 0.523-0.657 & 54.251 $\pm$ 1.874 $\pm$ 0.638\\
21 & 0.657-0.827 & 25.906 $\pm$ 1.162 $\pm$ 0.259\\
22 & 0.827-1.041 & 12.306 $\pm$ 0.730 $\pm$ 0.261\\
23 & 1.041-1.309 & 5.197 $\pm$ 0.439 $\pm$ 0.174\\
24 & 1.309-1.640 & 2.536 $\pm$ 0.279 $\pm$ 0.085\\
25 & 1.640-2.049 & 1.263 $\pm$ 0.206 $\pm$ 0.086\\
26 & 2.049-2.547 & 0.503 $\pm$ 0.113 $\pm$ 0.038\\
27 & 2.547-3.151 & 0.350 $\pm$ 0.104 $\pm$ 0.022\\
28 & 3.151-3.878 & 0.209 $\pm$ 0.063 $\pm$ 0.014\\
29 & 3.878-4.749 & 0.069 $\pm$ 0.042 $\pm$ 0.012\\

   \hline \hline
   \end{tabular}
   \label{Table:results_dimu_y2}
   \end{table}


\begin{thebibliography}{99}
\bibitem{coordinates}
Rapidity is defined by
$y = (1/2)\ln\left[\left(E-p_z\right)/\left(E+p_z\right)\right],$
where $E$ is the energy and $p_z$ is the momentum component parallel
to the proton beam axis..
Pseudorapidity is defined by 
\mbox{$\eta = -\ln\left[\tan(\theta/2)\right]$},
where $\theta$ is the polar angle.
\end{thebibliography}
\end{document}